\documentclass[prl,amsmath,amssymb,showpacs,superscriptaddress,floatfix,twocolumn]{revtex4}
\usepackage{graphicx}
\usepackage{bm}
\usepackage{lineno,hyperref}

\usepackage{epsf}
\usepackage{epstopdf}
\begin{document}
\title{Phase diagram of a two-dimensional system which stabilizes Kagome lattice}

\author{Yu. D. Fomin \footnote{Corresponding author: fomin314@mail.ru}}
\affiliation{Vereshchagin Institute of High Pressure Physics, Russian Academy of Sciences,
Kaluzhskoe shosse, 14, Troitsk, Moscow, 108840 Russia}
\affiliation{Moscow Institute of Physics and Technology (National Research University), 9 Institutskiy Lane, Dolgoprudny, Moscow region, 141701, Russia}

\date{\today}

\begin{abstract}
Phase diagram of a two-dimensional system with a potential which stabilizes Kagome lattice
is calculated. It is shown that this system demonstrate a set of crystalline and the
regions of stability of these phases are calculated. The scenarios of melting
of triangular and square crystals of the system are determined.

\end{abstract}

\pacs{61.20.Gy, 61.20.Ne}

\maketitle

\section{Introduction}

Two-dimensional (2D) systems are of great interest
for many fundamental and technological issues. They demonstrate
some unusual features which make them strongly different from the three-dimensional
(3D) ones. First of all, 2D systems do not have a genuine crystalline phase. As
it was pointed out by Pierls and Landau \cite{landau} and then by Mermin \cite{mermin}, 2D crystals do not
have long-range translational order. The translational order in 2D crystals
is quasi long-ranged, i.e. the correlation functions decay algebraically. At
the same time the orientational ordering of 2D crystals is long-ranged one,
like in the case of 3D space.

This difference between 2D and 3D crystals leads to difference in the melting mechanisms.
While in 3D melting is the first order phase
transition, at least three different scenarios of melting of 2d
crystals are widely discussed at the moment (see Ref. \cite{ufn} for review of the problem).
First, a 2D crystal can melt via the first order phase transition, as a 3D crystals do. The second
scenario is a so-called Berezinskii-Kosterlitz-Thouless-Halperin-Nelson-Young (BKTHNY) melting
\cite{bkt1,bkt2,bkt3,bkt4,halpnel1,halpnel2,halpnel3}.
In this scenario, melting proceeds via two stages. At the first step the orientational order
transforms from long-range to quasi-long range, while the translational order becomes short range.
It results in a phase which is neither solid, not isotropic liquid, since it demonstrates orientational
ordering. This phase is called hexatic phase. At the second step of the melting process the hexatic
phase looses the quasi-long-range orientational order and transforms into isotropic liquid.
Importantly, both transitions in BKTHNY scenario are continuous infinite order transitions, i.e.,
all derivatives of the free energy are continuous at these transitions. Finally, one more scenario
was proposed recently \cite{3a,3b,3c}. According to this scenario, melting of a 2D crystal proceeds
via two transitions, but while the transition from crystal to hexatic phase is a continuous transitions,
the hexatic phase transforms into the isotropic liquid by the first order phase transition. Below we
refer to this melting scenario as 'the third scenario'.

From the discussion above we see that melting of 2D crystals is a complex problem.
At the same time the variability of crystalline structures in 2D space is much lesser then in 3D one:
until recently it was supposed that 2D systems form a triangular crystalline structure only. The first
study which attracted a lot of attention to formation of other type of crystalline structures in 2D space
is the discovery of graphene \cite{grapene}. Later on more complex structures were discovered in 2D and quasi-2D systems, for instance,
square ice is formed when water is confined between two
graphene planes \cite{geim}, the square phase of 2D iron was discovered
in a slit pore with graphene walls \cite{iron}. Many complex structures
were observed in a system of colloidal particles in a magnetic field in the
work \cite{dobnikar}, However, up to now experimental observation of non-triangular
2D crystals is rather rare.

Surprisingly, many 2D systems which demonstrate non-triangular crystalline structures
are obtained in computer simulation. The publications which report different types
of crystalline and quasi-crystalline structures are numerous (see, for instance,
\cite{hzmiller,hzch,hzwe,we1,we2,we3,we4,we5,trusket2,trusket3,trusket4,qc1,qc2,qc3}
and references therein). Because of this one can expect that further experimental
studies will led us to find more experimental 2D systems with complex
phases.

A particular class of the models, which can demonstrate complex phase diagrams in the so-called
'core-softened systems', which are characterized by softening of the repulsive part of the interaction
potential \cite{deb}.
Depending on the shape of the potential core-softened systems can demonstrate complex phase diagrams (see Ref. \cite{3d1}
for review of the works on the phase diagrams of the core-softened systems). In our recent work a model core-softened system (Repulsive Shoulder System (RSS))
was proposed \cite{we-init}
\begin{equation}
  U(r)/\varepsilon= \left( \frac{d}{r} \right)^n+0.5 \left( 1 - tanh(k(r- \sigma))
  \right),
\end{equation}
where $n=14$, $k=10$ and the parameter $\sigma$ determines the
width of the repulsive shoulder of the potential. The phase diagram of RSS
in both 3D and 2D was studied (see \cite{we-init,s135b,s135c,s135d,s135e}  for 3D and \cite{we1,we2,we3,we4,we5}
for the 2D phase diagrams). It was shown that in both 3D and 2D this system
is characterized by very complex phase diagrams. In particular, in 2D case
the system with $\sigma_1/ \sigma=1.35$ demonstrates a square crystal and a dodecagonal
quasicrystalline phase \cite{we5}.

A generalization of this system was proposed in Ref. \cite{we-attract}. By adding
an attractive well to the potential (SRS - attractive well system - SRS-AW) we
obtained a system with a potential with both repulsive step and attractive well:
\begin{equation}\label{tanh2}
  U(r)/\varepsilon= \left( \frac{d}{r} \right)^n+ \sum_{i=1}^2 \left( 1 - tanh(k_i(r- \sigma_i))
  \right).
\end{equation}

In Refs.
\cite{we-attract,we-attract1,we-attract2,we-attract3} we investigated the
phase diagram of 3D SRS-AW system at different parameters of the potential.

The results of exploration of both SRS and SRS-AW systems showed that
the phase diagrams of these systems are very sensitive to the parameters
of the potential. Because of this other groups of researchers decided to
use this potential to an inverse problem: given a particular structure they aim
find parameters of SRS-AW potential which can stabilize it. In Ref.
\cite{str2d3d} a method to solve this problem was proposed. The authors
proposed a scheme to compare the chemical potentials of different crystalline
structures at zero temperature and given pressure and employed this scheme
to find a potential (not of SRS-AW form) which stabilize a diamond structure.
The same method was applied to the SRS-AW potential to stabilize different
3D and 2D structures in the works \cite{en1,en2,en3}. In particular,
in Ref. \cite{en3} a parameterizations of the SRS-AW potential which
stabilizes the Kagome lattice was proposed. Later on the properties of this
particular system were investigated in Ref. \cite{rice-kgm}. All phases
presented in the system were identified. However, the complete phase diagram
of has not been constructed.

The goal of the present paper is to calculate the phase diagram of 2D
SRS-AW system with the parameters which stabilize Kagome lattice.
We find all structures existing in the system and calculate the regions of
their stability in the density-temperature plane. We also discuss the
scenarios of melting of the low-density triangular phase and
the square phase of the system.

\section{Systems and methods}

In this work we investigate the SRS-AW system (Eq. \ref{tanh2})
with parametrization which stabilizes Kagome lattice
\cite{en3}. In order to employ this parametrization, we rewrite
the potential in the form of Ref. \cite{en3}:

\begin{equation}\label{tanh3}
  U(r)/\varepsilon= A \left( \frac{\sigma}{r} \right)^n+ \sum_{i=1}^2 \lambda_i \left( 1 - tanh(k_i(r/ \sigma- d_i))
  \right) + U_{shift} 
\end{equation}
where $U_{shift}=Pr^2+Qr+R$ is used to make both the potential and
its first and second derivatives continuous at cut-off distance
$r_c$. The parameters of the potential are given in Table I. The parameters
$\sigma$ and $\varepsilon$ are used as the units of length and energy. All other
quantities are expressed basing on these parameters. Only these dimensionless units
are used throughout the paper.

\begin{table}
\begin{tabular}{|c|c|}
  \hline
  A & 0.01978 \\
  n & 5.49978 \\
  $\lambda_1$ & -0.06066 \\
  $k_1$ & 2.53278 \\
  $d_1$ & 1.94071 \\
  $\lambda_2$ & 1.06271 \\
  $k_2$ & 1.73321 \\
  $d_2$ & 1.04372 \\
  $r_c$ & 3.0 \\
  $P$ & 0.007379 \\
  $Q$ & 0.04986 \\
  $R$ & -0.085054 \\
  \hline
\end{tabular}
\caption{The potential parameters \ref{tanh3} used in the present
study.}
\end{table}

We study the behavior of the system by means of molecular dynamics method. The system is
simulated in a rectangular (or square depending on the phase simulated) box with periodic boundary conditions.
We follow the same methodology that was used in Refs. \cite{hzwe,hz72}, i.e. firstly we study a small system
of 4000 particles in a wide range of densities in order to get a rough estimation of the phase diagram.
After that we simulate a larger system of 20000 particles in order to establish the melting scenarios
of the low-density triangular phase and 22500 particles for the square phase. In all cases the system was simulated for $5 \cdot 10^7$ steps
with the time step $dt=0.001$. First $2 \cdot 10^7$ steps were discarder from the analysis and used
for equilibration of the system. At the last $3 \cdot 10^7$ the properties of the system were calculated.
We calculated the equations of state (EOS), i.e. the dependence of pressure on density at fixed temperature.
The structure of the system was probed by radial distribution functions (RDFs) $g(r)$ and diffraction
patterns (DP). The latter were calculated as $S(\bf{k})= <
\frac{1}{N} \left( \sum_i^Ncos(\bf{kr}_i)\right)^2+\left(
\sum_i^Nsin(\bf{kr}_i)\right)^2>$.

RDFs and DPs give enough information to find out the structure of the system. However, they do not
allow to establish the melting scenarios. In the present work we combine several methods to make
a justified conclusion on the melting scenario. Firstly, we monitor the EOS of the system. If
a first order transition takes place in the system, the EOS demonstrates a Mayer-Wood loop.
Indeed, such a loop can appear even in the case of the second order phase transition \cite{ising}. However,
we are not aware of any melting scenario which involves a second order transition. The BKT-type transitions
which appear in BKTHNY and in the third scenarios are of infinite order, and no loop on the
equation of state can appear in this case. Because of this we consider the presence of a loop
as an evidence of the presence of the first order phase transition \cite{hz72}.

We also analyze the orientational and translational order parameters of the system.
The local orientational order parameter (OOP) is defined as \cite{halpnel1,halpnel2,dfrt5,dfrt6}:

\begin{equation}
\psi_6({\bf r_i})=\frac{1}{n(i)}\sum_{j=1}^{n(i)} e^{i
n\theta_{ij}}\label{psi6loc},
\end{equation}
where $\theta_{ij}$ is the angle of the vector between particles
$i$ and $j$ with respect to the reference axis. The sum over
$j$ counts $n(i)$ nearest-neighbors of $j$. Voronoi construction is used to find
the nearest neighbors of a particle.

The global OOP is obtained by averaging of the local OOP over the whole system:
\begin{equation}
\Psi_6=\frac{1}{N}\left<\left|\sum_i \psi_6({\bf
r}_i)\right|\right>.\label{psi6}
\end{equation}

The translational order parameter is defined as \cite{halpnel1, halpnel2, dfrt5, dfrt6}:
\begin{equation}
\Psi_T=\frac{1}{N}\left<\left|\sum_i e^{i{\bf G
r}_i}\right|\right>, \label{psit}
\end{equation}
where ${\bf r}_i$ is the position  of $i$-tj particle and {\bf
G} is the reciprocal-lattice vector of the first shell of the
crystal lattice.

Further analysis of ordering in the system is undertaken by calculations
of correlation functions of OOP and TOP. The orientational correlation function (OCF) is calculated as:
\begin{equation}
g_6(r)=\frac{\left<\Psi_6({\bf r})\Psi_6^*({\bf 0})\right>}{g(r)},
\label{g6}
\end{equation}
where $g(r)=<\delta({\bf r}_i)\delta({\bf r}_j)>$  is RDF of the system.
In the case of crystal the OCF has a flat shape, i.e. it does not decay at the
scale of the sample. In the hexatic phase the long-range
behavior of $g_6(r)$ has the form $g_6(r)\propto r^{-\eta_6}$ with
$\eta_6 \leq \frac{1}{4}$ \cite{halpnel1, halpnel2}. In isotropic liquid
OCF decays exponentially.

The translational correlation function (TCF) is defined as
\begin{equation}
g _T(r)=\frac{<\exp(i{\bf G}({\bf r}_i-{\bf r}_j))>}{g(r)},
\label{GT}
\end{equation}
where $r=|{\bf r}_i-{\bf r}_j|$. In the solid phase the long-range
behavior of $G_T(r)$ has the form $g_T(r)\propto r^{-\eta_T}$ with
$\eta_T \leq \frac{1}{3}$, i.e. it decays algebraically \cite{halpnel1, halpnel2}. In the
hexatic phase and isotropic liquid $g_T$ decays exponentially, i.e.
these phases demonstrate short-range translational order only.

The combination of EOS with analysis of order parameters and their correlation functions
allows to distinguish between the three melting scenarios unambiguously.

\section{Results and Discussion}

\begin{figure}
\includegraphics[width=8cm,height=8cm]{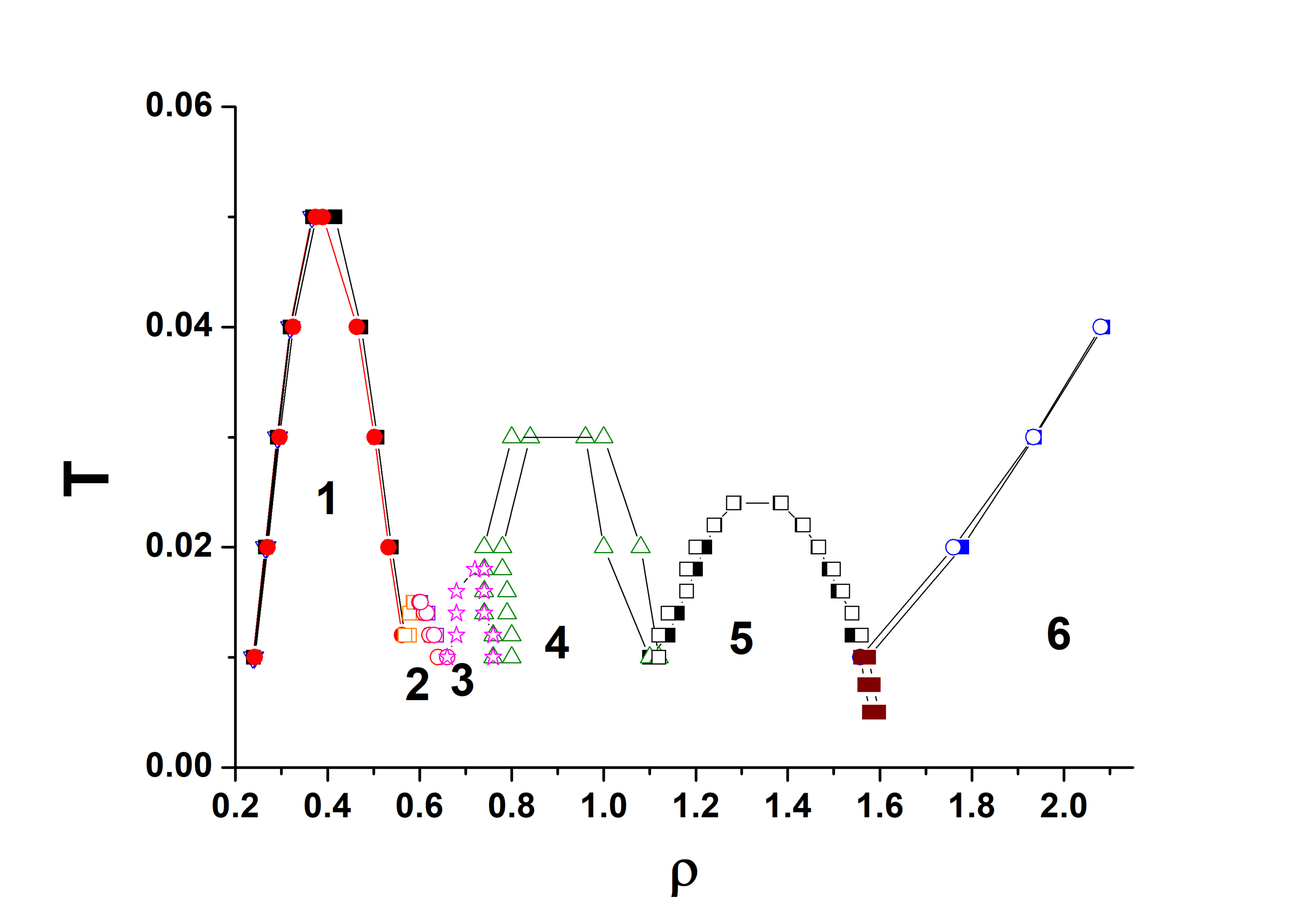}%

\caption{\label{pd} Phase diagram of the system. The phases are denoted by numbers. 1 - low-density
triangular phase, 2 - square phase, 3 - dimers (pairs), 4 - oblique crystal (stripe phase), 5 - Kagome
lattice, 6 - high-density triangular crystal.}
\end{figure}

In this section we calculate the phase diagram of the system under investigation. As it has been
mentioned above, all phases presented in the system have been identified in Ref. \cite{rice-kgm}, but
the authors have not calculated the complete regions of stability of these phases. Going from low density
to high one, one observes the following sequence of phases: fluid $\rightarrow$ hexatic $\rightarrow$
low density triangular solid $\rightarrow$ hexatic $\rightarrow$ square solid (possibly with the advent of tetratic phase)
$\rightarrow$ dimers (or pairs in terms of Ref. \cite{rice-kgm}) $\rightarrow$ oblique crystal (stripe phase)
$\rightarrow$ Kagome lattice $\rightarrow$ high density triangular crystal. Fig. \ref{pd} shows
the whole phase diagram obtained in the present study. Below we discuss the properties of all phases and the transition
lines in more detail.

The first crystal structure which appears in the system is the low-density triangular phase. As the density
increases it transforms into the square crystal. We will discuss the properties of these phases
in more detail below.

\begin{figure}
\includegraphics[width=8cm,height=6cm]{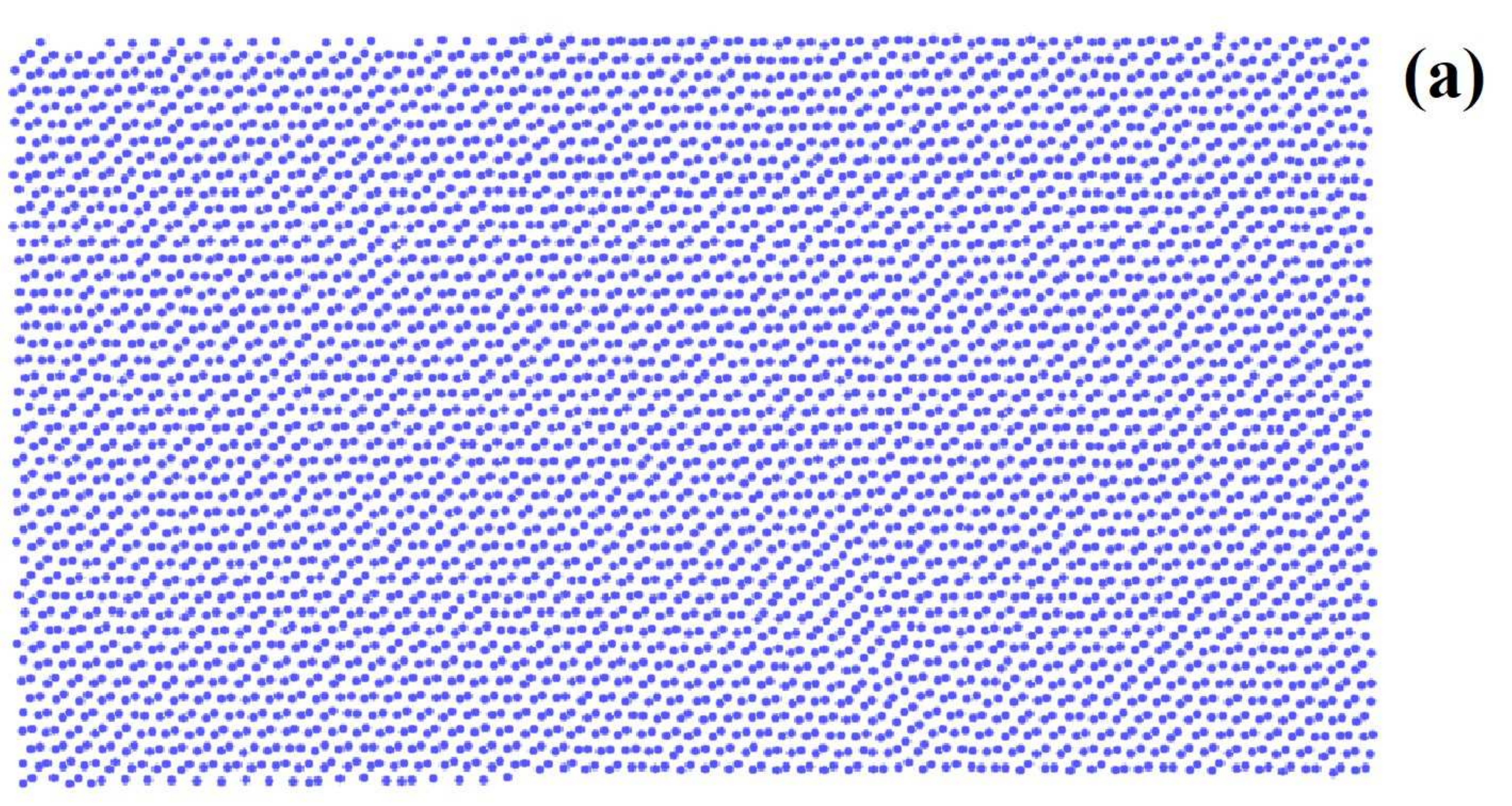}%

\includegraphics[width=8cm,height=6cm]{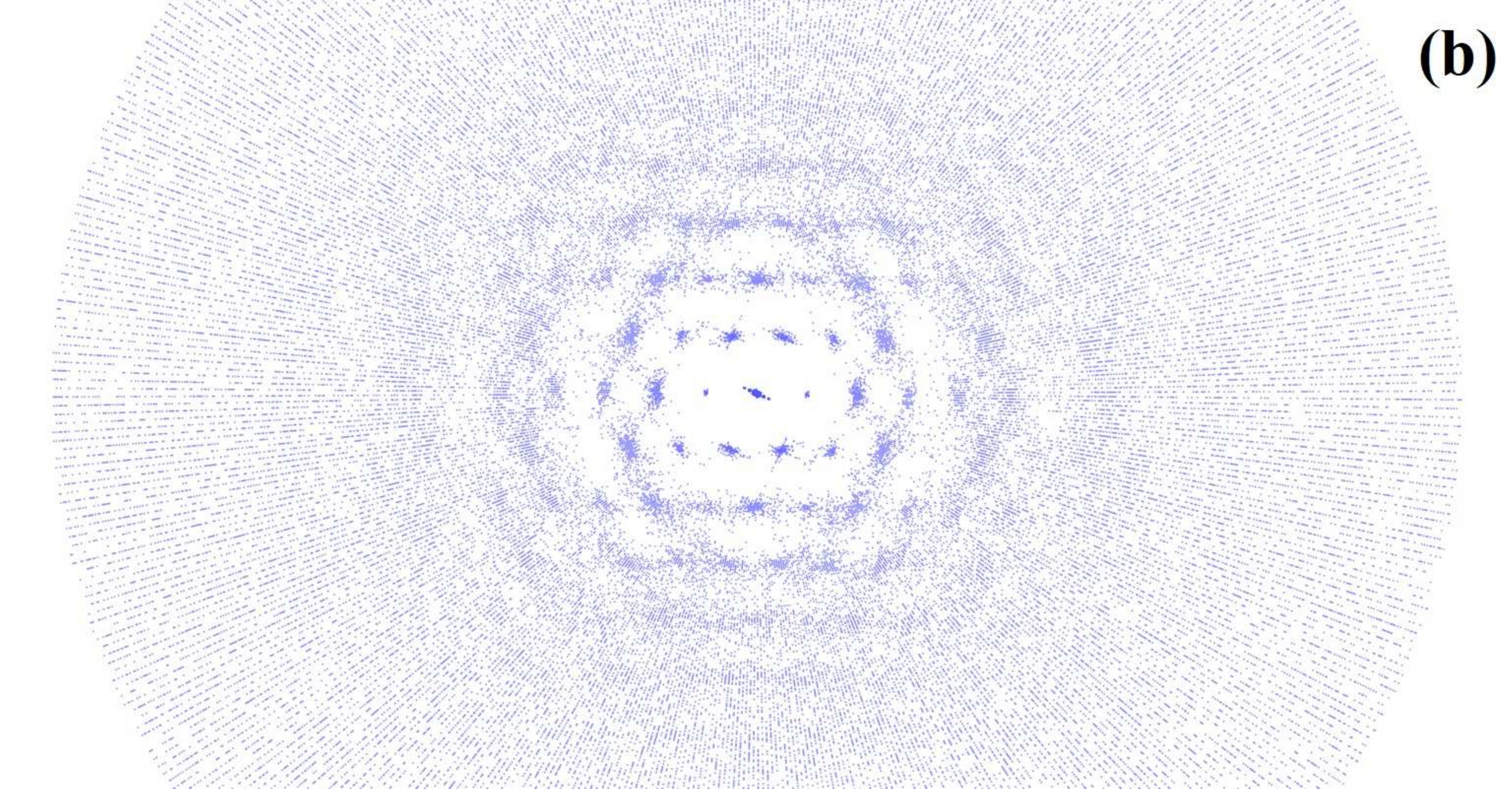}%

\caption{\label{r07} (a) A snaphot of the system at $T=0.012$ and $\rho=0.7$ (phase of dimers). (b) Diffraction
pattern at the same point.}
\end{figure}

The next structure after the square crystal is a phase of dimers (or pairs). The mechanisms
of formation of this phase are discussed in Ref. \cite{rice-kgm}. It is also shown in the
same publication that the centers of mass of the dimers form a stretched hexagonal lattice.
Fig. \ref{r07} (a) and (b) show a snapshot of the phase of dimers and a diffraction pattern of
the same phase. The diffraction pattern shows clear hexagonal features, which confirms the
conclusions of Ref. \cite{rice-kgm}.

\begin{figure}
\includegraphics[width=8cm,height=8cm]{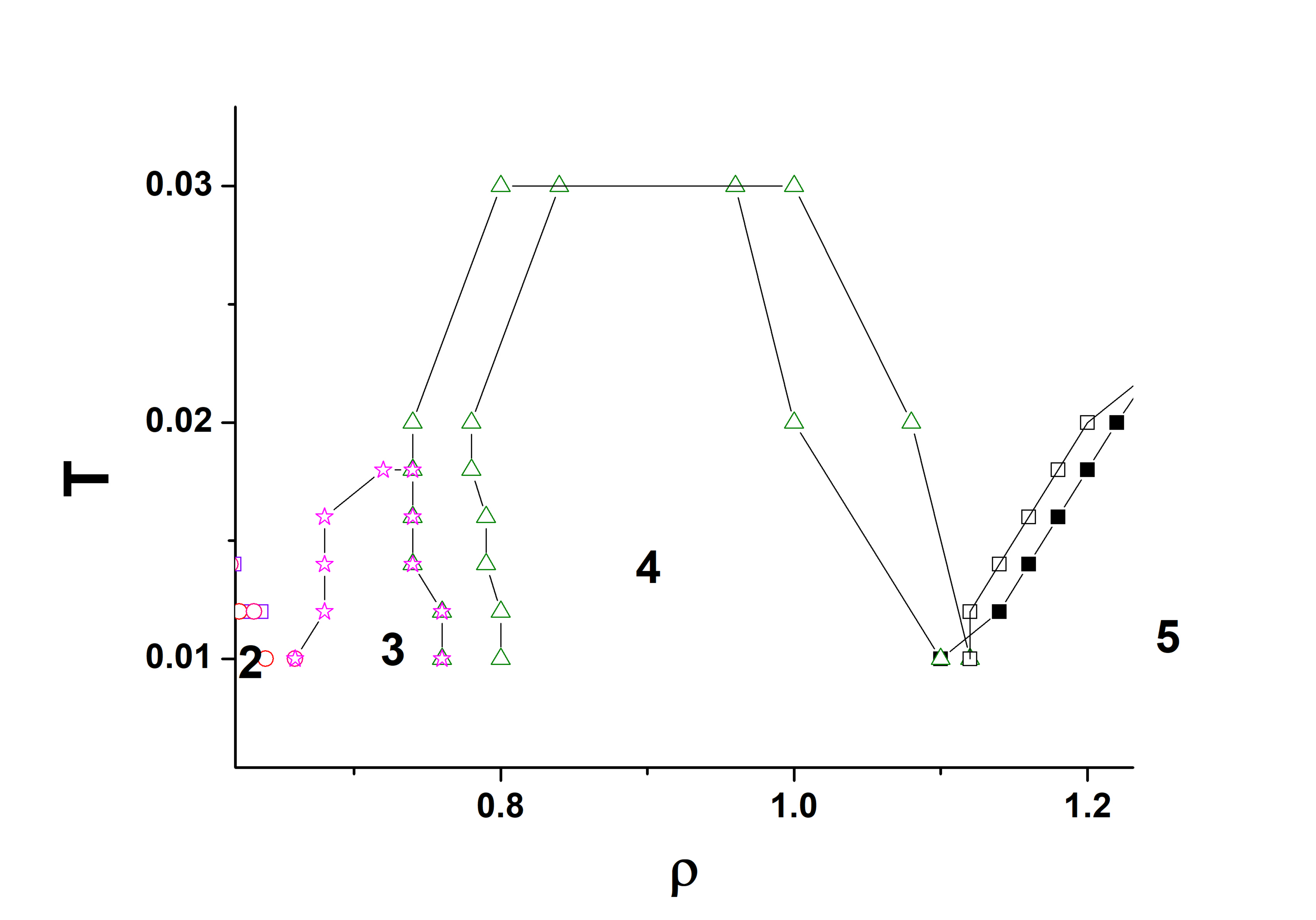}%

\caption{\label{dimers-str} The phase diagram of the system in the vicinity of the
phase of dimers (3) and stripes (4).}
\end{figure}

When the density increases the phase of dimers transforms into stripe phase. The structure of the stripe phase
was identified in our recent publication \cite{stripe}: the stripe phase is an oblique
crystal. The primitive vectors of this phase are of different length and the angle between
these vectors is about 40 degrees. The mechanisms of transformation of the dimers into stripes
were revealed in Ref. \cite{rice-kgm}. Since dimers and stripes were widely discussed in the
recent papers \cite{rice-kgm,stripe} we do not discuss these phases in detail here, but give the
limits of stability of these phases only (Fig. \ref{dimers-str}). Interestingly, the phase of
dimers is stable at low temperatures only. The oblique crystal is stable up to much higher temperatures.
Because of this it coexists with dimers at low temperature and with liquid at the higher one.

At higher densities the stripe phase transforms into Kagome lattice which on further
densification experiences a transition into high-density triangular phase. The Kagome phase
is stable in a rather wide range of densities. Although Kagome lattice is very unusual
for one component systems, it can be observed in complex colloidal systems \cite{kgm-exp}.
In this respect it is of special interest to find out the mechanisms responsible for the formation
of the Kagome lattice. However, this problem is far from being solved.

\begin{figure}
\includegraphics[width=6cm,height=6cm]{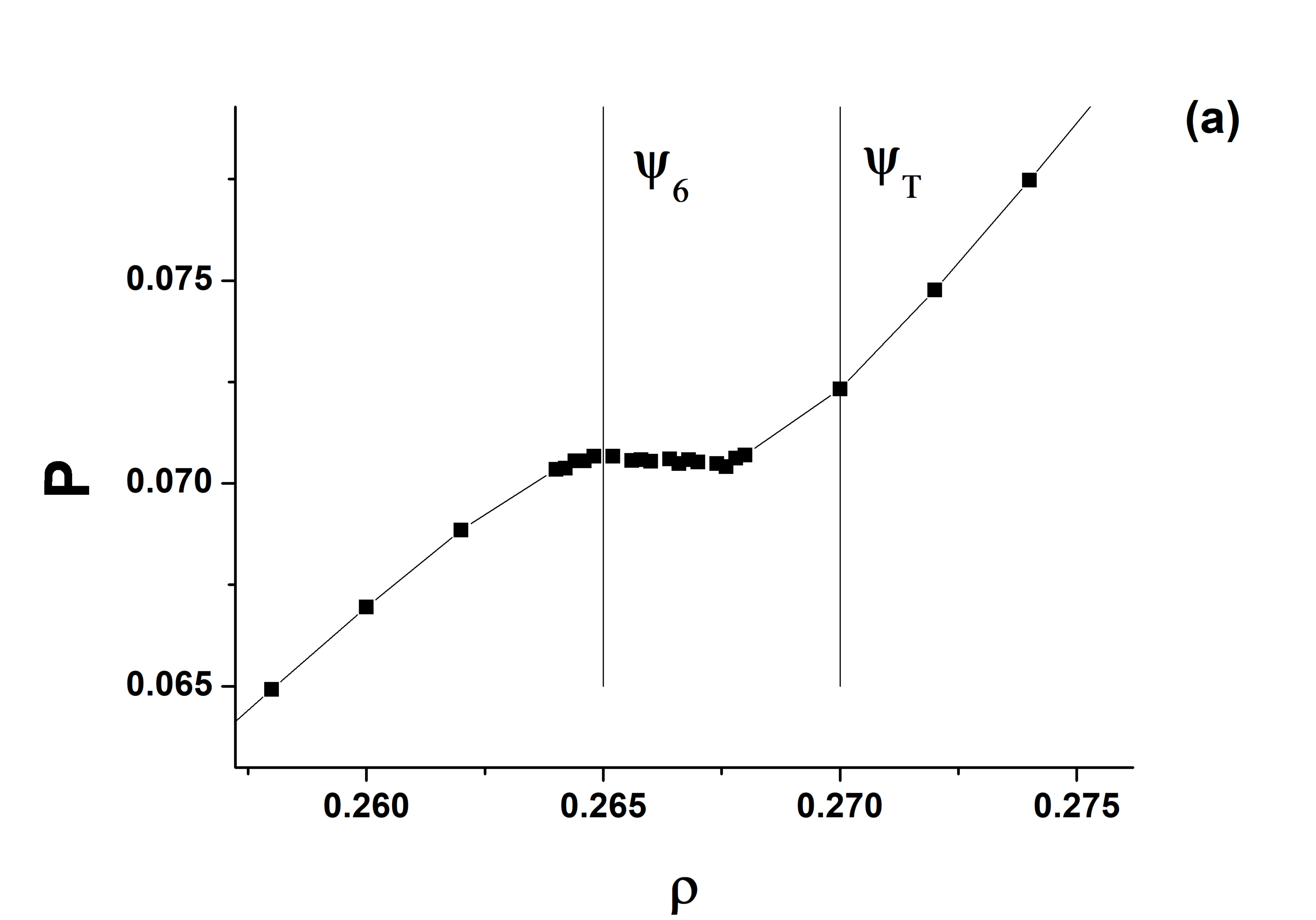}%

\includegraphics[width=6cm,height=6cm]{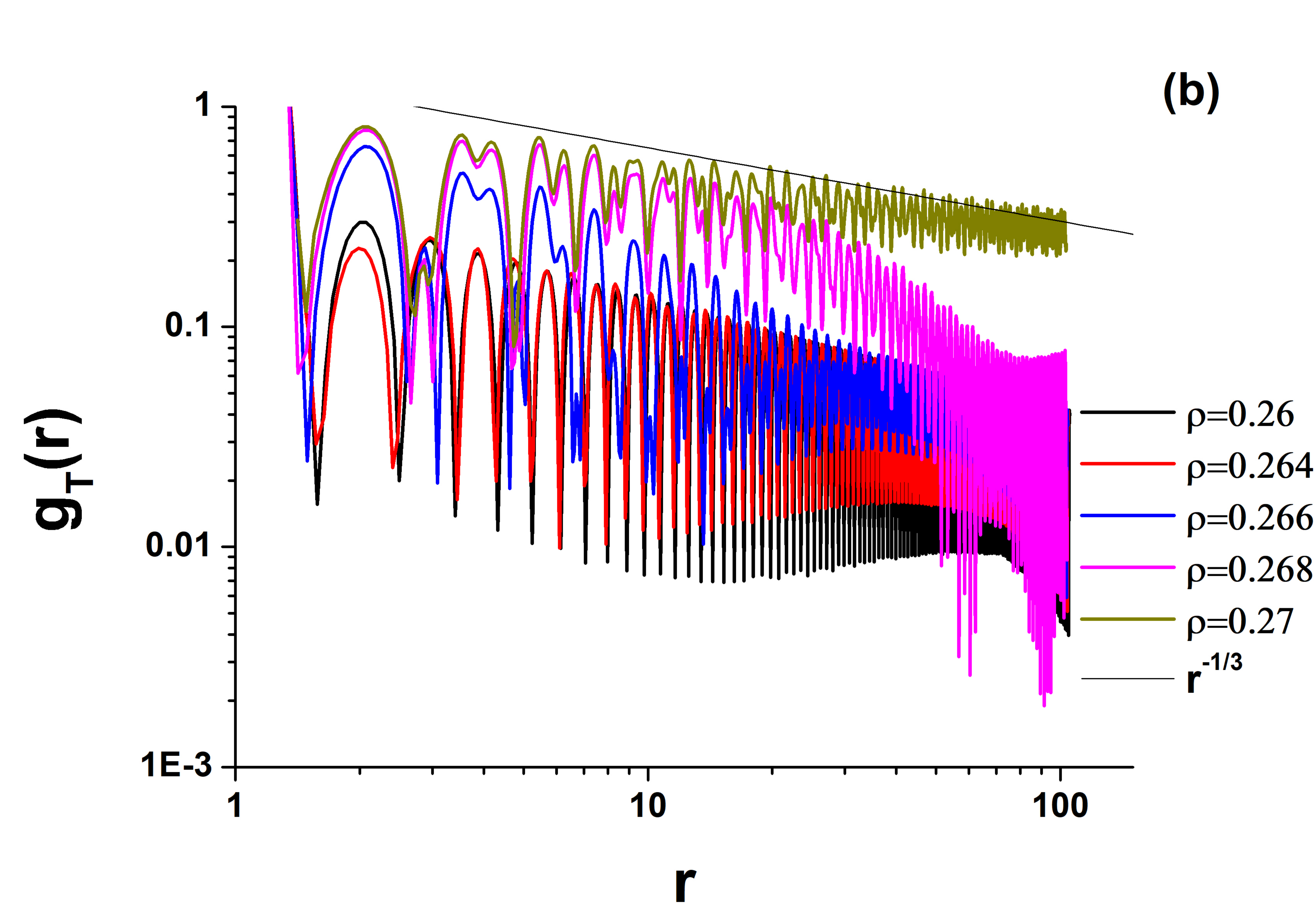}%

\includegraphics[width=6cm,height=6cm]{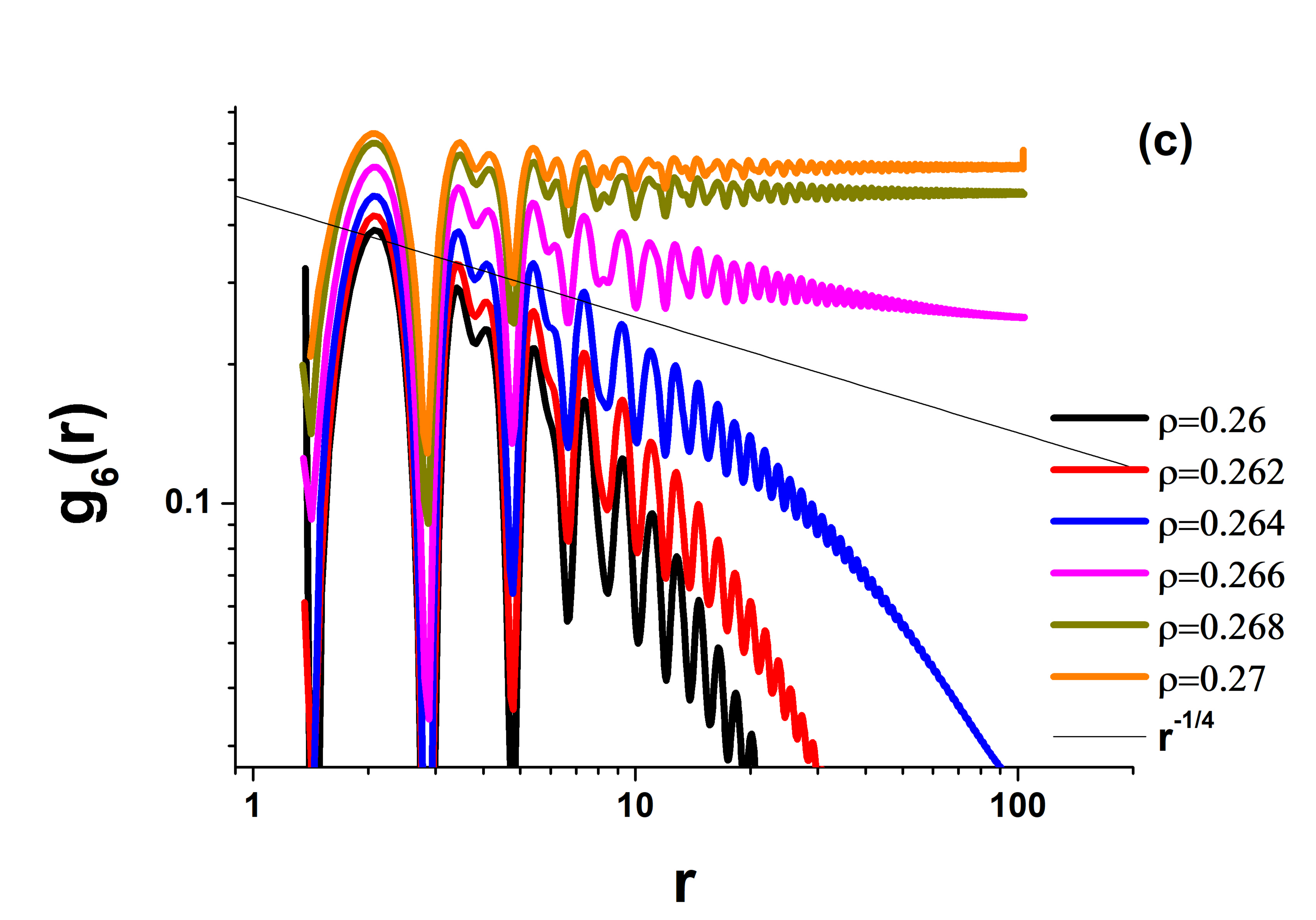}%

\caption{\label{ld-tr-left} (a) Equation of state of the system at $T=0.02$ in the region of the low-density
branch of the melting line of the low-density triangular crystal. (b) TCF in the same region. (c) OCF in the same region.
The vertical lines at panel (a) show the limit of stability of the crystal with respect to the hexatic phase ($\psi_T$) and
the limit of stability of the hexatic phase with respect to the isotropic liquids ($\psi_6$).}
\end{figure}

Having identified all structures in the system, we turn to more detailed investigation of
melting lines of the low-density triangular and square phases.
First we consider the low-density triangular phase. This phase is studied with the system of 20000 particles. Fig. \ref{ld-tr-left}
shows the equation of state, TCF and OCF of this system at the low-density branch of the melting line at $T=0.02$.
One can see that the EOS demonstrates a Mayer-Wood loop, i.e., first order transition takes place in the system. In order to distinguish
between the first and the third scenarios we calculate the orientational and translational correlations functions.
From these plots one can see that, if one goes from higher densities to the lower ones, the crystal looses its
stability to the hexatic phase prior to the Mayer-Wood loop, i.e., there is a continuous transition from
crystal to the hexatic phase, and, therefore, the loop corresponds to the first-order transition between the hexatic phase
and isotropic liquid. Therefore, the third scenarios is realized in this case.

\begin{figure}
\includegraphics[width=6cm,height=6cm]{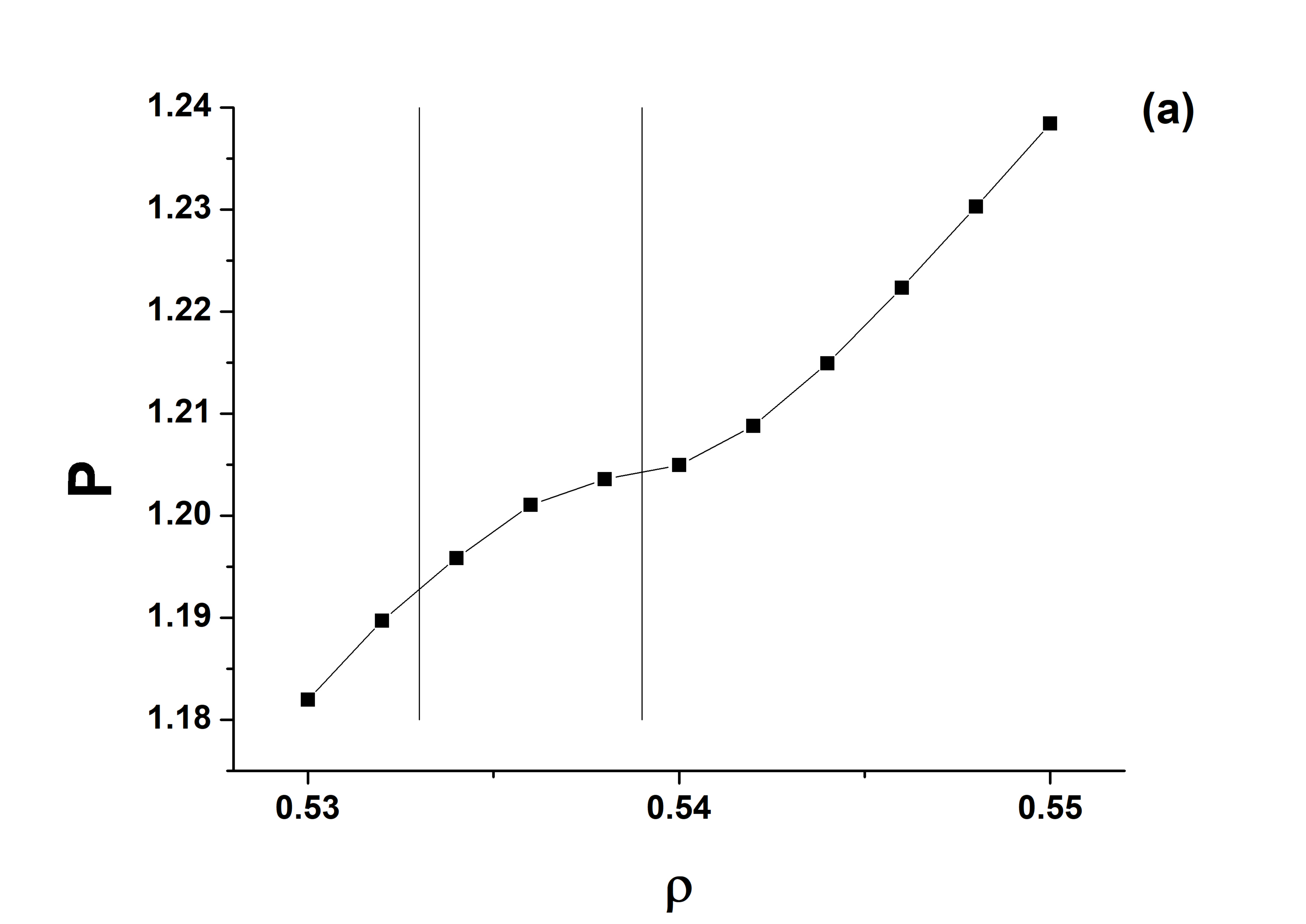}%

\includegraphics[width=6cm,height=6cm]{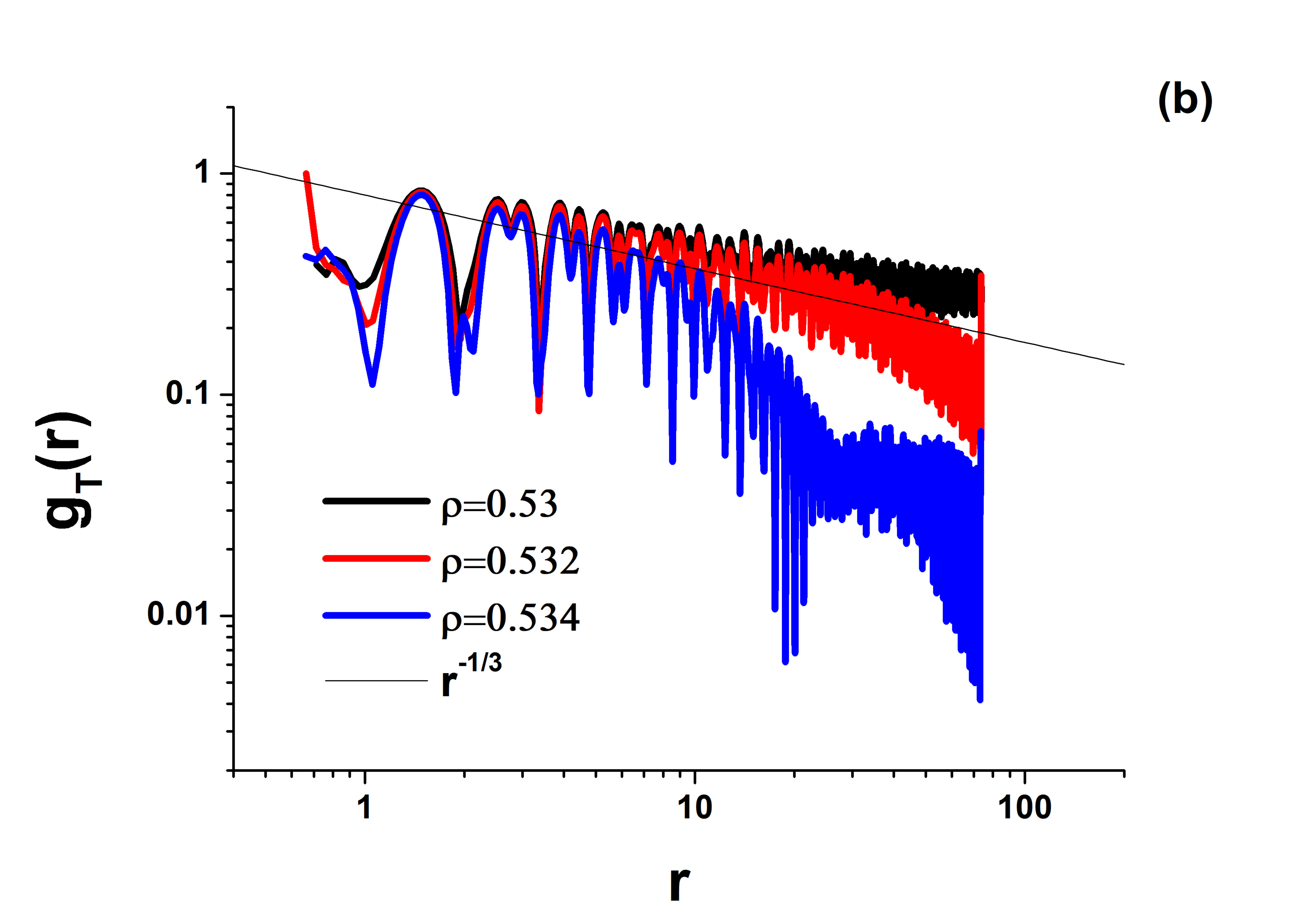}%

\includegraphics[width=6cm,height=6cm]{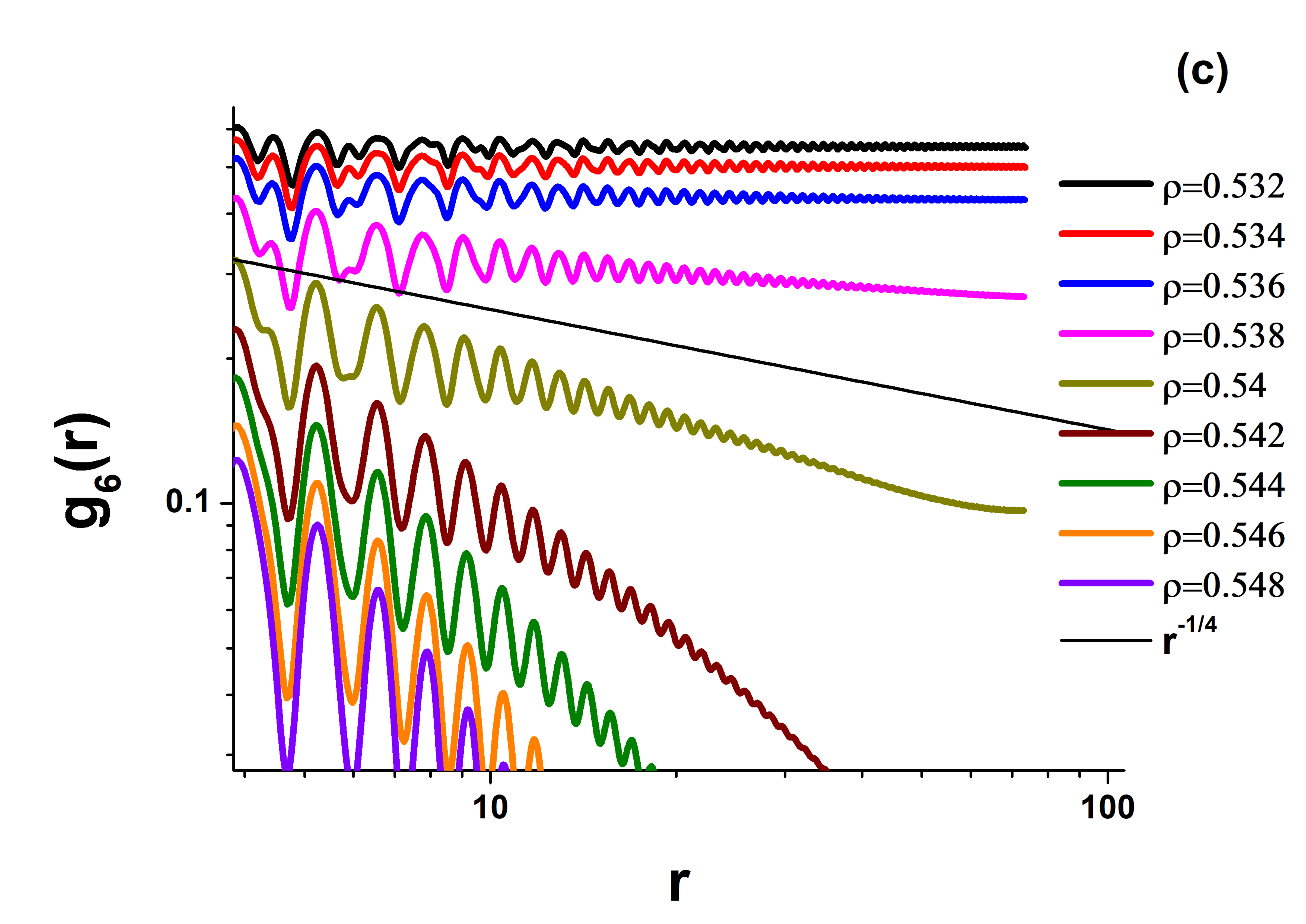}%

\caption{\label{ld-tr-right} (a) Equation of state of the system at $T=0.02$ in the region of the high-density
branch of the melting line of the low-density triangular crystal. (b) TCF in the same region. (c) OCF in the same region.
The vertical lines at panel (a) show the limit of stability of the crystal with respect to the hexatic phase ($\psi_T$) and
the limit of stability of the hexatic phase with respect to the isotropic liquids ($\psi_6$).}
\end{figure}

The behavior of the melting line at the high-density branch of the melting line of the low-density
triangular crystal is different. In this case we do not observe the Mayer-Wood loop (Fig. \ref{ld-tr-right}), and therefore
the melting appears via BKTHNY scenario. The limits of stability of the crystal and hexatic phase are
calculated from the TCF and OCF functions.

\begin{figure}
\includegraphics[width=8cm,height=8cm]{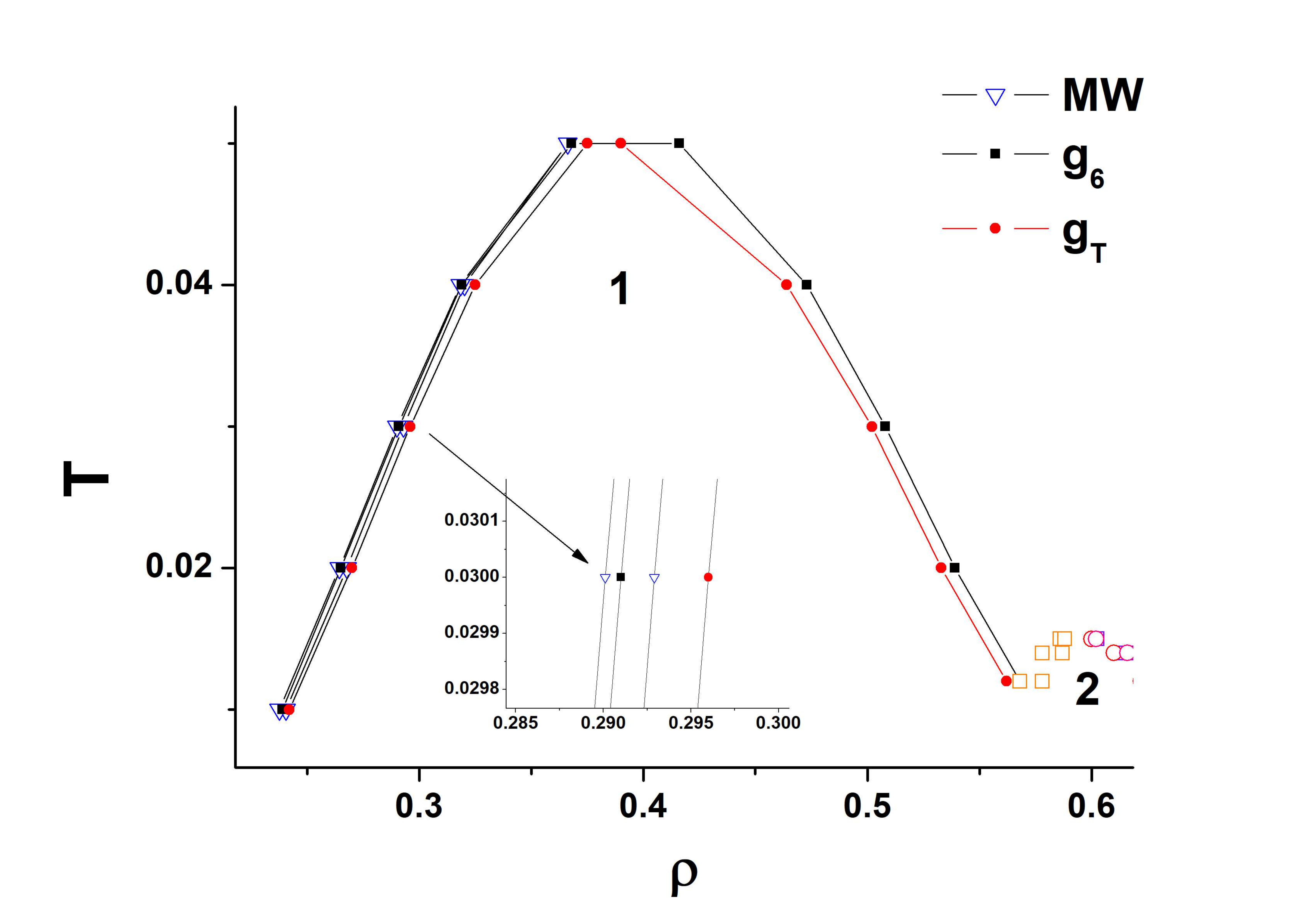}%

\caption{\label{ld-tr} A part of the phase diagram in the vicinity of the low-density triangular
phase. The line MW means the melting points obtained from the Mayer-Wood loop. the lines $g_6$
and $g_T$ mark the limits of stability of the hexatic phase with respect to liquid and
the crystal with respect to the hexatic phase obtained from OCF and TCF respectively. The
inset enlarges the left branch at $T=0.03$. From this inset one can see that going from
the higher density to the lower through the melting line firstly the crystal transforms into hexatic phase
and then the hexatic phase transforms into liquid through the first order transition, because
the point of the $g_6$ line is in between of the MW line points.}
\end{figure}

The phase diagram in the vicinity of the low-density triangular phase is shown in Fig. \ref{ld-tr}.

In our previous publications we studied the Hertz system, which is characterized by the interaction potential
$U(r)=\varepsilon (r-\sigma)^{\alpha}$, if $r<\sigma$ and zero otherwise. We found that in the case of $\alpha=5/2$
there are two tricritical points on the melting line of the low-density triangular phase \cite{hzwe}. One of them
is at the maximum of the melting line and another one is on the high-density branch. In the case of
$\alpha =7/2$ a single tricritical point at the low density branch is observed \cite{hz72}.
The low-density triangular phase of the RSS-AW system melts via the third scenario
at the low-density branch and the BKTHNY scenario at the high-density branch. However, we do not find the second tricritical point in the RSS-AW system.
Therefore, the melting behavior of the low-density triangular phase of SRR-AW system is similar, but
not equivalent to the one of the Hertzian spheres with $\alpha=5/2$.

\begin{figure}
\includegraphics[width=6cm,height=6cm]{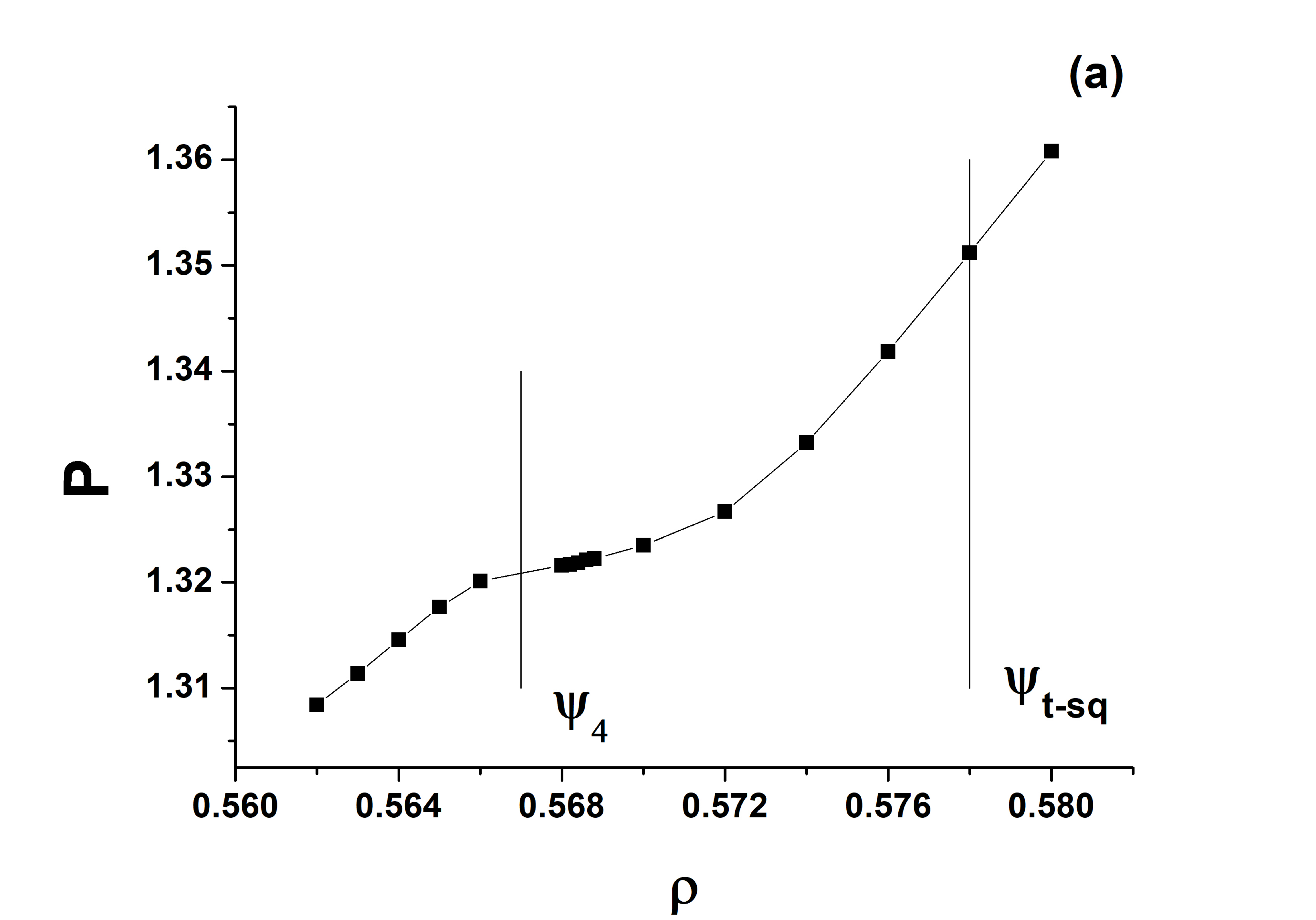}%

\includegraphics[width=6cm,height=6cm]{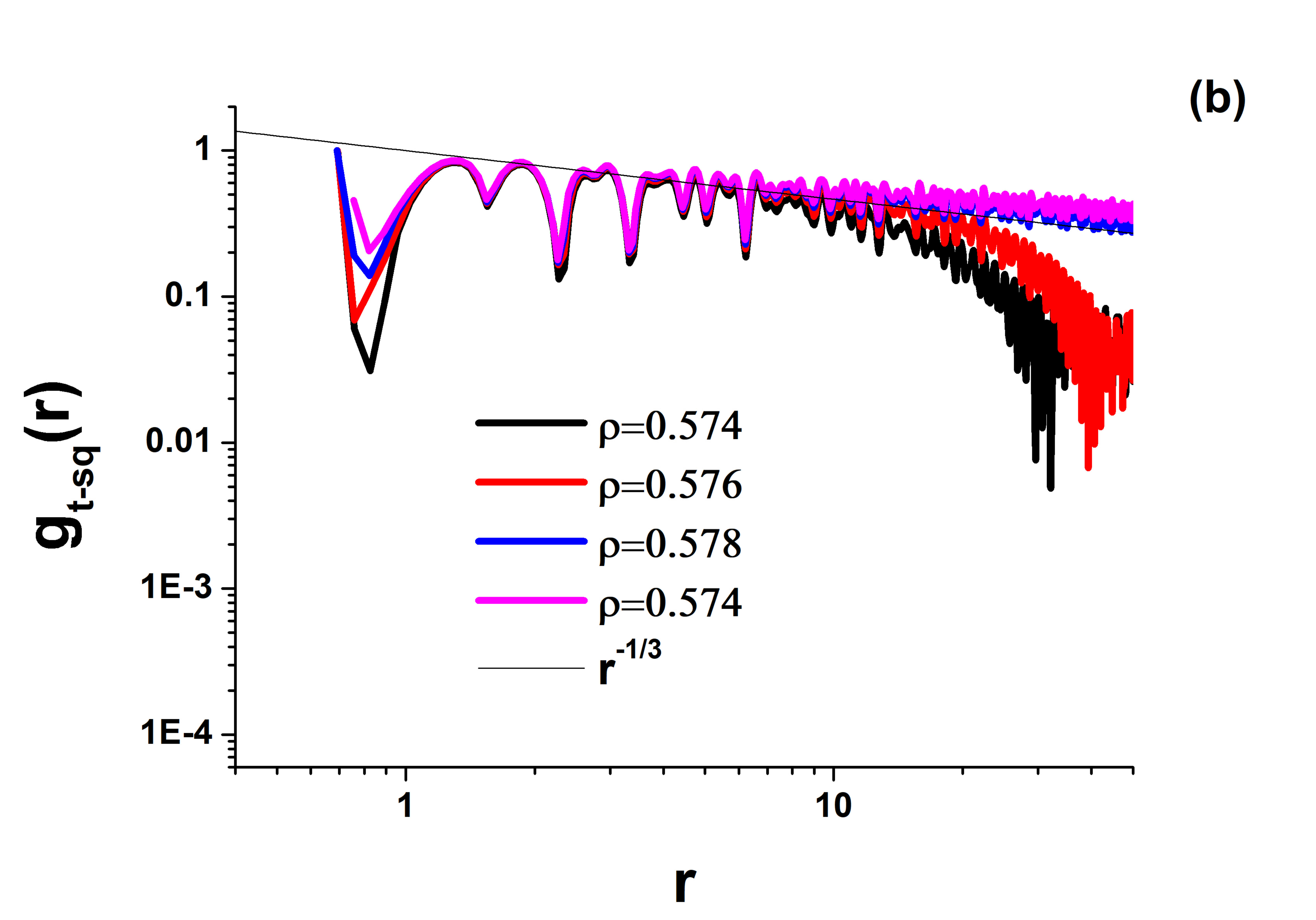}%

\includegraphics[width=6cm,height=6cm]{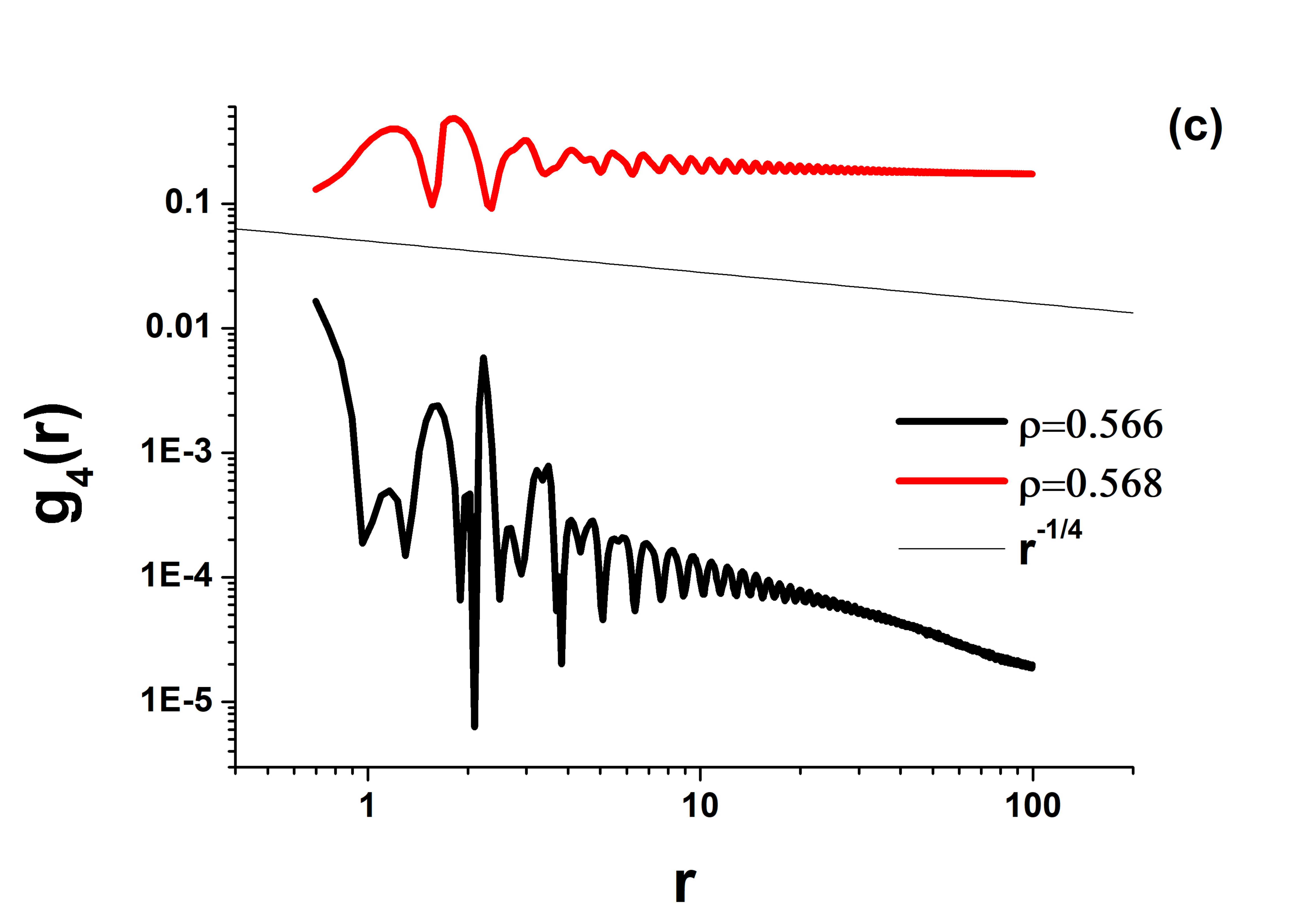}%

\caption{\label{sq-left} (a) Equation of state of the system at $T=0.012$ in the region of the low-density
branch of the melting line of the square crystal. (b) TCF in the same region. (c) OCF in the same region.
The vertical lines at panel (a) show the limit of stability of the crystal with respect to the tetratic phase ($\psi_{t-sq}$) and
the limit of stability of the tetratic phase with respect to the isotropic liquids ($\psi_4$).}
\end{figure}

The same calculations are performed for the square phase. A system of 22500 particles in a square box is used in this case.
Fig. \ref{sq-left} shows the behavior of the system across the melting line of the square phase at $T=0.012$. One can see
that there is no Mayer-Wood loop on the equation of state. At the same time the $g_{t-sq}$ and $g_4$ criteria give
different transition points. Because of this we conclude that the system melts via BKTHNY scenario.

\begin{figure}
\includegraphics[width=6cm,height=6cm]{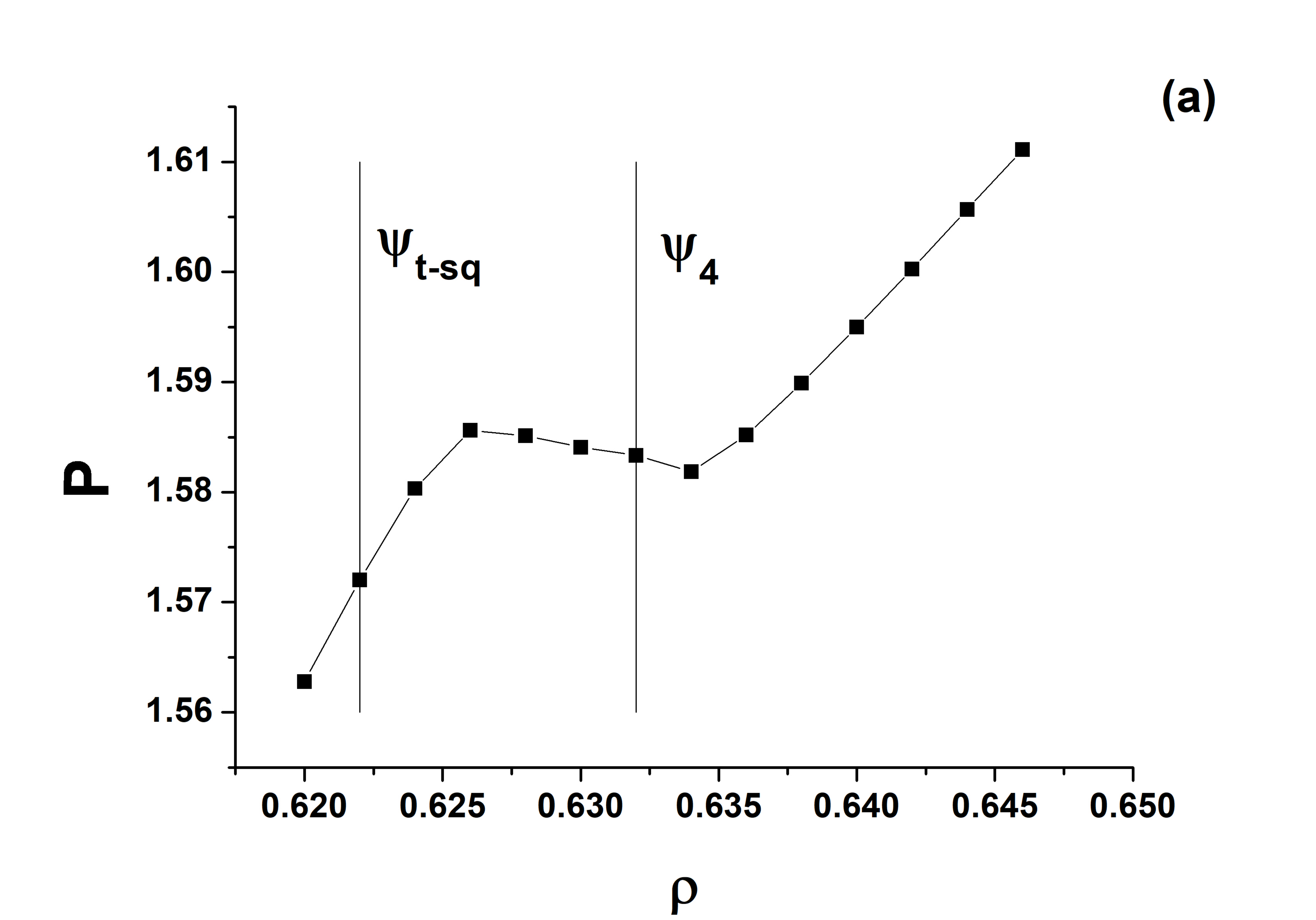}%

\includegraphics[width=6cm,height=6cm]{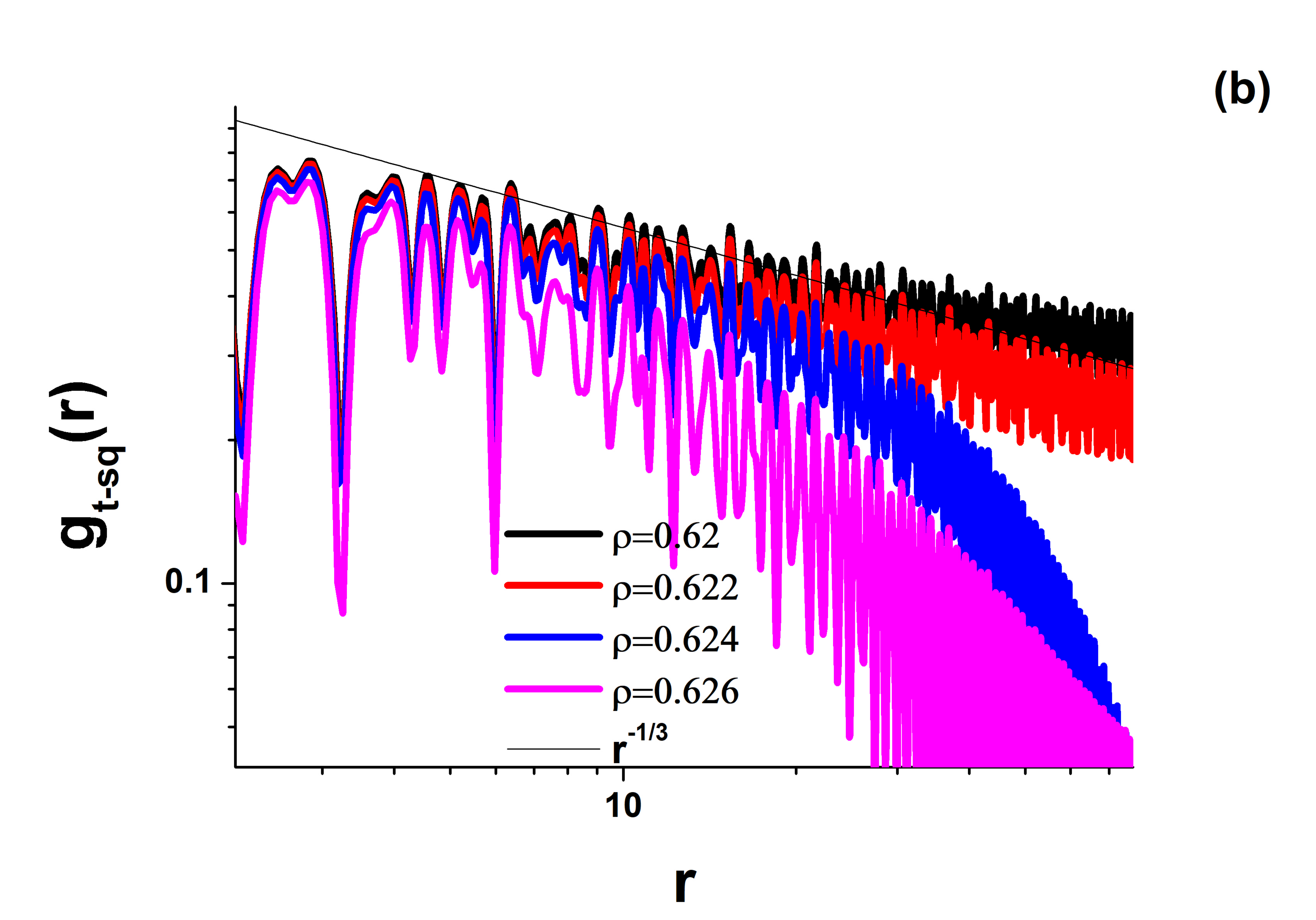}%

\includegraphics[width=6cm,height=6cm]{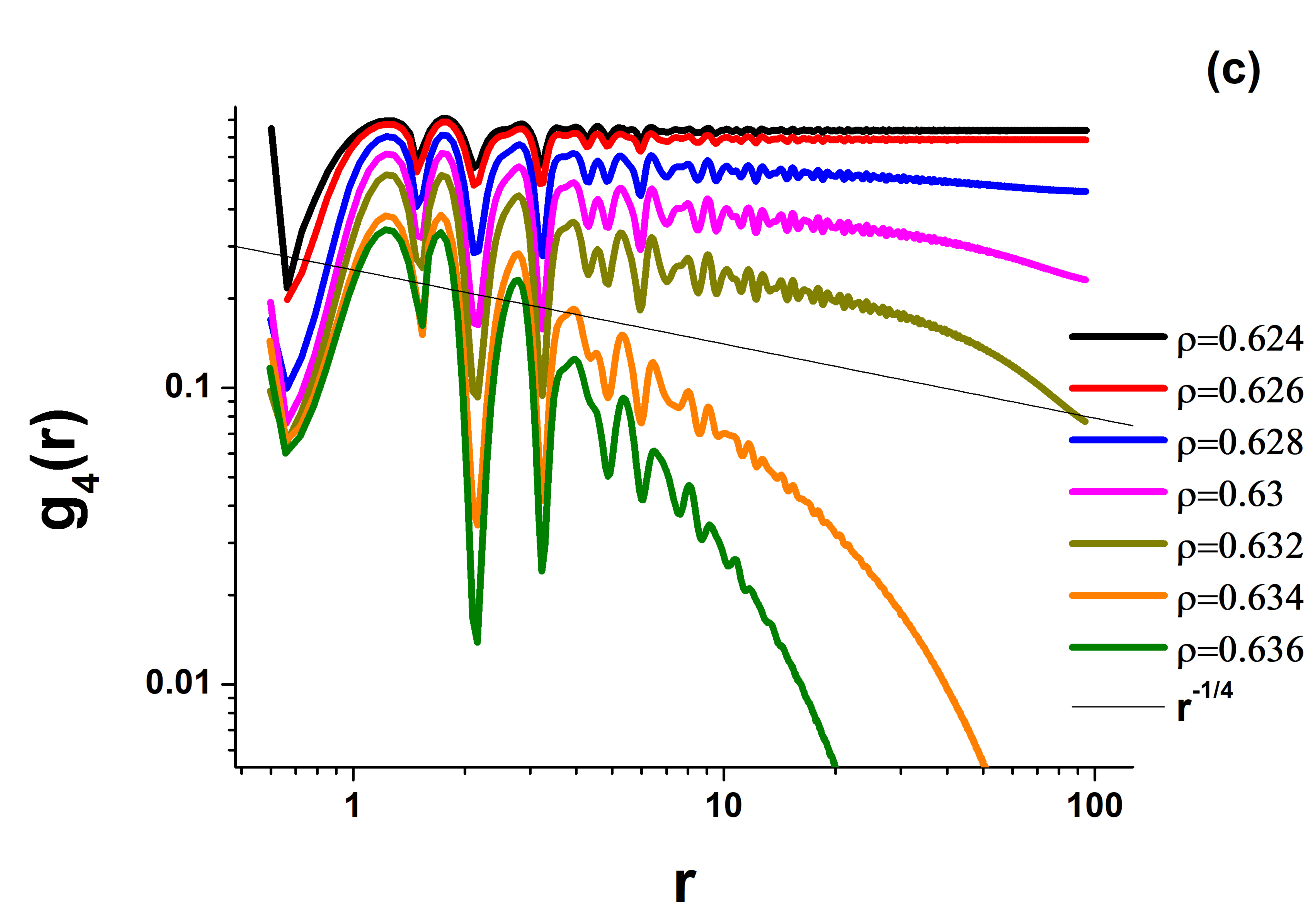}%

\caption{\label{sq-right} (a) Equation of state of the system at $T=0.012$ in the region of the high-density
branch of the melting line of the square crystal. (b) TCF in the same region. (c) OCF in the same region.
The vertical lines at panel (a) show the limit of stability of the crystal with respect to the tetratic phase ($\psi_{t-sq}$) and
the limit of stability of the tetratic phase with respect to the isotropic liquids ($\psi_4$).}
\end{figure}

However, the high-density branch of the square crystal behaves differently (Fig. \ref{sq-right}). First of all, there is
a clear Mayer-Wood loop, therefore, first order phase transition takes place in the system. The limit of stability of the crystal
with respect to the tetratic phase preempts the first order transition points, while the limit of stability
of the tetratic phase with respect to the liquid obtained on the basis of OCF criterion is inside the loop. Therefore,
the third scenario takes place in the system.

\begin{figure}
\includegraphics[width=8cm,height=8cm]{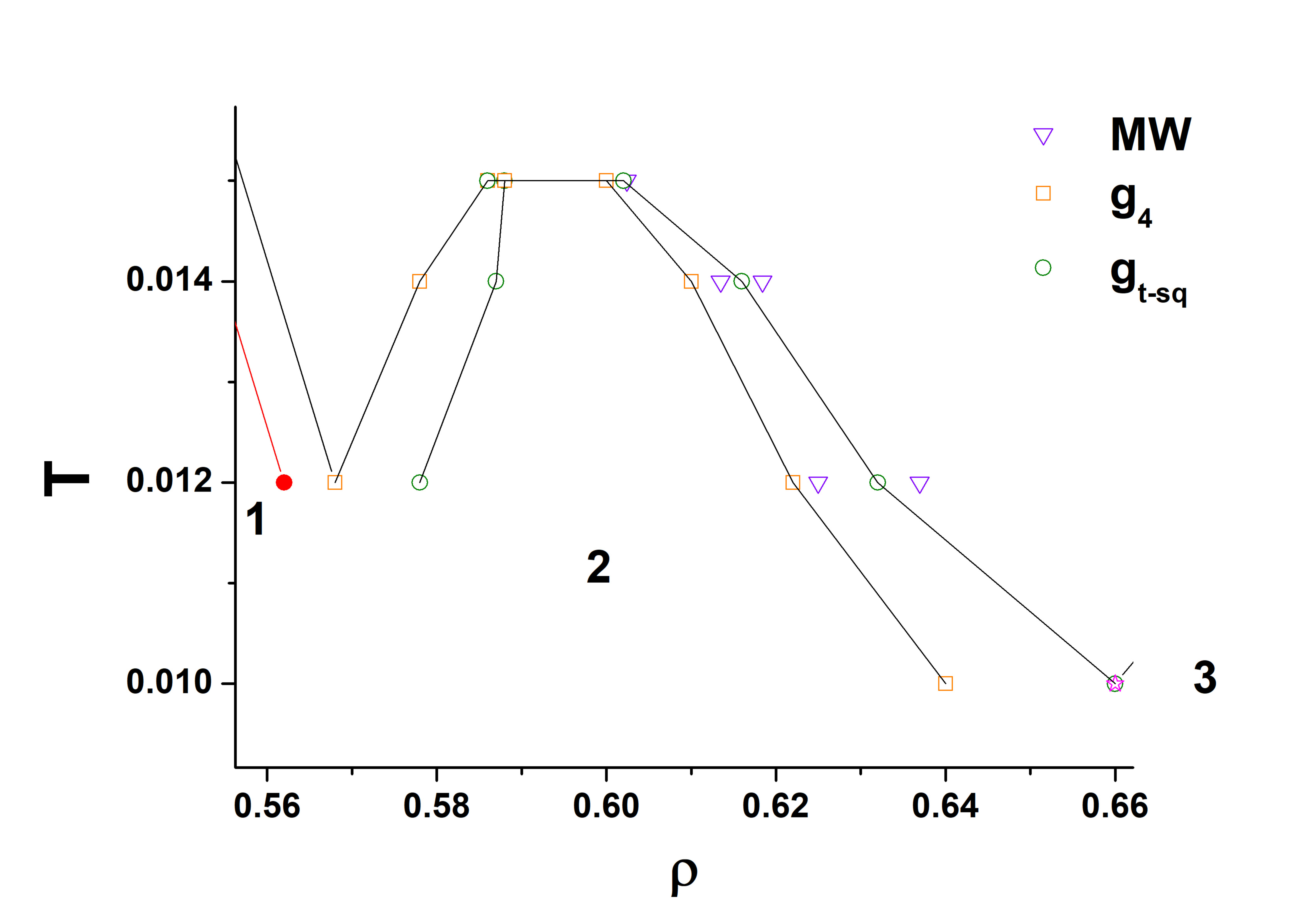}%

\caption{\label{pd-sq} A part of the phase diagram in the vicinity of the square crystal.
The line MW means the melting points obtained from the Mayer-Wood loop. The lines $g_4$
and $g_{t-sq}$ mark the limits of stability of the hexatic phase with respect to liquid and
the crystal with respect to the hexatic phase obtained from OCF and TCF respectively.}
\end{figure}

We conclude that the melting line of the square crystal consists of BKTHNY part at the low-density branch
and the third scenario at the high density one. The tricritical point is at the maximum of the melting line.
The part of the phase diagram in the vicinity of the square crystal is shown in Fig. \ref{pd-sq}.

\section{Conclusions}

In conclusion, we have studied the phase diagram of the SRS-AW system with parametrization
which stabilizes Kagome lattice. We confirmed the sequence of phases which was found
in Ref. \cite{rice-kgm} and for the first time calculated the complete phase diagram.
We studied in details the melting line of the low-density triangular and square phases.
Both of these phases demonstrate two different melting scenarios with tricritical
points at the maximum of the melting line. In the case of the low-denisty
triangular phase the low-density branch of the melting line proceeds by the third scenario,
while the high density one - via BKTHNY scenario. In the case of the square
phase the situation is reversed: the low density branch proceeds via BKTHNY scenario,
while the high-density one by the third scenario.

\section{Acknowledgments}
This work was carried out using computing
resources of the federal collective usage centre "Complex for
simulation and data processing for mega-science facilities" at NRC
"Kurchatov Institute", http://ckp.nrcki.ru, and supercomputers at
Joint Supercomputer Center of the Russian Academy of Sciences
(JSCC RAS). The work was supported by the Russian Science
Foundation (Grant No 19-12-00092).


\begin{thebibliography}{99}


\bibitem{landau} L.D. Landau, E.M. Lifshitz, Statistical Physics, Pergamon Press, Ltd., London, 1958, p. 482.

\bibitem{mermin} N.D. Mermin, Phys. Rev. 176 (1968) 250.



\bibitem{ufn} V. N. Ryzhov, E. E. Tareyeva, Yu. D. Fomin, and E. N.
Tsiok, Phys. Usp. 60 857–885 (2017).

\bibitem{bkt1} V.L. Berezinskii, Zh. Eksp. Teor. Fiz. 59 (1970) 907; Sov. Phys.—JETP 32 (1970) 493.
\bibitem{bkt2} V.L. Berezinskii, Zh. Eksp. Teor. Fiz. 61 (1971) 1144; Sov. Phys.—JETP 34 (1971) 610.
\bibitem{bkt3} J.M. Kosterlitz, D.J. Thouless, J. Phys. C 5 (1972) L124.
\bibitem{bkt4} J.M. Kosterlitz, D.J. Thouless, J. Phys. C 6 (1973) 1181.
\bibitem{halpnel1} B. I. Halperin and D. R. Nelson, {\it Phys. Rev. Lett.}, 1978, {\bf
41}, 121.
\bibitem{halpnel2} D. R.Nelson and B. I. Halperin, {\it Phys. Rev. B: Condens.
Matter Mater. Phys.}, 1979, {\bf 19}, 2457.

\bibitem{halpnel3} A. P. Young, {\it Phys. Rev. B: Condens.
Matter Mater. Phys.}, 1979, {\bf 19}, 1855.




\bibitem{3a} E. P. Bernard and W. Krauth, Phys. Rev. Lett. 107, 155704 (2011).

\bibitem{3b} M. Engel, J. A. Anderson, Sh. C. Glotzer, M. Isobe, E. P. Bernard, and W. Krauth,
Phys. Rev. E 87, 042134 (2013).

\bibitem{3c} S. C. Kapfer and W. Krauth, Phys. Rev. Lett. 114, 035702 (2015).






\bibitem{grapene} A. K. Geim and K. S. Novoselov, Nature Materials 6, 183-191 (2007).

\bibitem{geim} G. Algara-Siller, O. Lehtinen, F. C. Wang, R. R. Nair, U. Kaiser, H. A.Wu, A. K. Geim and I. V. Grigorieva, Nature  519, 443
(2015).

\bibitem{iron} Jiong Zhao et al.,  Science 343, 1228 (2014).

\bibitem{dobnikar} N. Osterman, D. $Babi\check{c}$,1 I. Poberaj, J. Dobnikar, and P.
Ziherl, Phys. Rev. Lett. 99, 248301 (2007).

\bibitem{hzmiller} W.L. Miller and A. Cacciuto, Soft Matter 7, 7552 (2011).

\bibitem{hzch} M. Zu, P. Tan, and N. Xu, Nat. Comm. 8, 2089 (2017).

\bibitem{hzwe} Yu. D. Fomin, E. A. Gaiduk, E. N. Tsiok, and V. N.
Ryzhov, Mol. Phys. doi.org/10.1080/00268976.2018.1464676

\bibitem{we1} D.E. Dudalov, Yu.D. Fomin, E.N. Tsiok, and V.N.
Ryzhov, Journal of Physics: Conference Series 510 (2014) 012016.

\bibitem{we2} D. E. Dudalov, E. N. Tsiok, Yu. D. Fomin, and V. N.
Ryzhov, J. Chem. Phys. 141, 18C522 (2014).

\bibitem{we3} E. N. Tsiok, D. E. Dudalov, Yu. D. Fomin, and V. N.
Ryzhov, Phys. Rev. E 92, 032110 (2015).

\bibitem{we4} D.E. Dudalov, Yu.D. Fomin, E.N. Tsiok, and V.N.
Ryzhov, Soft Matter, 10, 4966 (2014).

\bibitem{we5} N. P. Kryuchkov, S. O. Yurchenko, Yu. D. Fomin, E. N. Tsiok, and V. N. Ryzhov, Soft Matter 14, 2152 (2018)


\bibitem{trusket2} A. Jain, J. R. Errington and Th. M. Truskett,
Phys. Rev. X 4, 031049 (2014).

\bibitem{trusket3} W. D. $Pi\tilde{n}eros$, M. Baldea, and Th. M.
Truskett, J. Chem. Phys. 144, 084502 (2016).

\bibitem{trusket4} W. D. $Pi\tilde{n}eros$, M. Baldea, and Th. M.
Truskett, J. Chem. Phys. 145, 054901 (2016).

\bibitem{qc1} M. Engel and H.-R. Trebin, Phys. Rev. Lett. 98, 225505
(2007).

\bibitem{qc2} T. Dotera, T. Oshiro, and P. Ziherl, Nature 506, 208–211 (2014).

\bibitem{qc3} H. Pattabhiraman and M. Dijkstra, J. Chem. Phys. 146, 114901
(2017).

\bibitem{deb} P. G. Debenedetti, V. S. Raghavan and S. S. Borik, J. of Phys. Chem. 95,
4540 (1991).




\bibitem{3d1} Ryzhov V N, Tareyeva E E, Fomin Yu D, Tsiok E N,
Complex phase diagrams of systems with isotropic potentials: results of computer simulation,
2020, Phys. Usp., $accepted; DOI: 10.3367/UFNe.2018.04.038417$

\bibitem{we-init} Yu. D. Fomin, N. V. Gribova, V. N. Ryzhov, S. M. Stishov, and Daan
Frenkel, J. Chem. Phys. 129, 064512 (2008).

\bibitem{s135b} Fomin Yu D, Tsiok E N, and Ryzhov V N, 2011, J. Chem. Phys. 135, 234502

\bibitem{s135c} Fomin Yu D and Ryzhov V N, 2011, Physics Letters A 375, 2181–2184

\bibitem{s135d} Fomin Yu D, Tsiok E N, and Ryzhov V N, 2013, Eur. Phys. J. Special Topics 216, 165–173

\bibitem{s135e} Fomin Yu D, Tsiok E N, and Ryzhov V. N, 2011, J. Chem. Phys. 135, 124512

\bibitem{we-attract} Yu. D. Fomin, E. N. Tsiok, and V. N. Ryzhov,
J. Chem. Phys. 134, 044523 (2011).

\bibitem{we-attract1} Yu. D. Fomin, E. N. Tsiok, and V. N. Ryzhov,
Phys. Rev. E 87, 042122 (2013).

\bibitem{we-attract2} Yu. D. Fomin, E. N. Tsiok, and V. N. Ryzhov,
Eur. Phys. J. Special Topics 216, 165–173 (2013).

\bibitem{we-attract3} Yu. D. Fomin, E. N. Tsiok, and V. N. Ryzhov,
J. Chem. Phys. 135, 124512 (2011).


\bibitem{str2d3d} E. Marcotte, F. H. Stillinger, and S.
Torquato, J. Chem. Phys. 134, 164105 (2011).

\bibitem{en1} A. Jain, J. R. Errington and Th. M. Truskett, Soft Matter, 9, 3866 (2013).

\bibitem{en2} W. D. Pineros, M. Baldea, and Th. M. Truskett, J. Chem. Phys. 144, 084502 (2016).

\bibitem{en3} W. D. Pineros, M. Baldea, and Th. M. Truskett, J. Chem. Phys. 145, 054901 (2016).

\bibitem{rice-kgm} L. Nowack and S. A. Rice, J. Chem. Phys. 151, 244504 (2019)

\bibitem{hz72} E. N. Tsiok, E. A. Gaiduk, Yu. D. Fomin and V. N. Ryzhov, Soft Matter,16, 3962-3972 (2020)

\bibitem{ising} Juan J. Alonso and Julio F. Fernandez, it Phys. Rev. E 59, 2659 (1999).



\bibitem{dfrt6} E. N. Tsiok, Y. D. Fomin, V. N. Ryzhov, {\it Physica A}, 2018, {\bf 490}, 819.


\bibitem{dfrt5}  E. N. Tsiok, D. E. Dudalov, Yu. D. Fomin, and V. N.
Ryzhov, {\it Phys. Rev. E: Stat. Phys., Plasmas, Fluids, Relat.
Interdiscip. Top.}, 2015, {\bf 92}, 032110.




\bibitem{lammps} http://lammps.sandia.gov/

\bibitem{stripe} Yu.D. Fomin, E.N. Tsiok, V.N. Ryzhov, Physica A 527, 121401 (2019).

\bibitem{kgm-exp} Q. Chen, S. Chul Bae, and S. Granick, Nature 469,
381–384 (2011).


\end{thebibliography}
\end{document}